\documentclass[journal=aamick,manuscript=article]{achemso}

\usepackage[version=3]{mhchem} % Formula subscripts using \ce{}
\usepackage[T1]{fontenc}       % Use modern font encodings
\usepackage{amsmath}
\usepackage{upgreek}
\author{Katy L. Dickinson}
\affiliation[UoE_SoPA]
{SUPA School of Physics \& Astronomy, The University of Edinburgh, Edinburgh, EH9 3FD, United Kingdom}
\author{Ulrich K. Wiegand}
\affiliation[UoE_SoBS]
{School of Biomedical Sciences, The University of Edinburgh, Edinburgh, EH8 9XD, United Kingdom}
\author{Job H. J. Thijssen}
\affiliation[UoE_SoPA]
{SUPA School of Physics \& Astronomy, The University of Edinburgh, Edinburgh, EH9 3FD, United Kingdom}
\email{j.h.j.thijssen@ed.ac.uk}
\phone{+44 (0)131 6505274}

\title{Soft meets hard -- how does freeze-thaw cycling affect the microstructure of particle-stabilised emulsions?}

\begin{document}

\begin{tocentry}
  \includegraphics{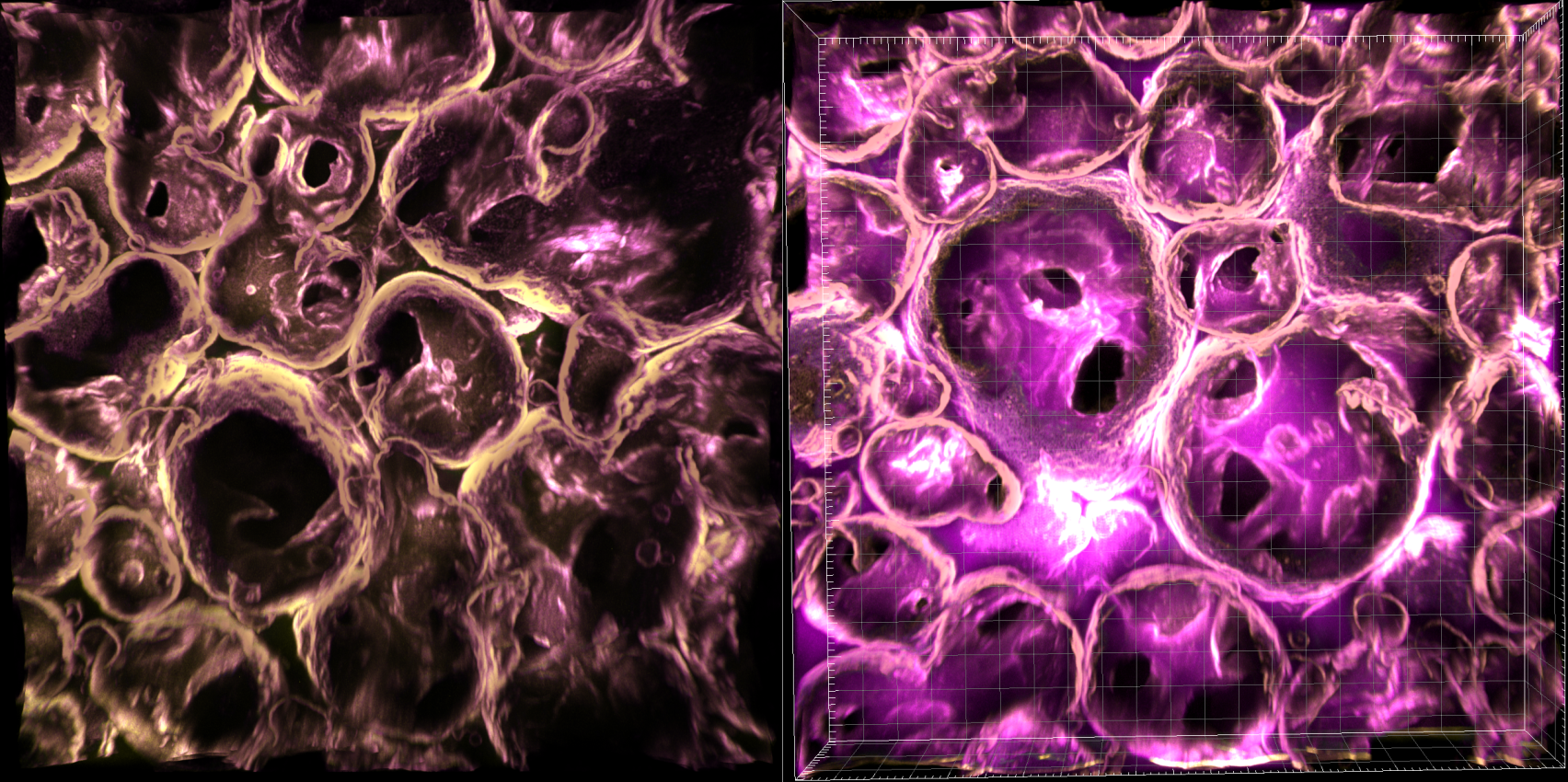}
\end{tocentry}

\begin{abstract}

The freeze-thaw cycling of particle-stabilised emulsions can alter the emulsion structure and stability. This could have significant consequences for using particle stabilisation in industrial applications where increased stability is generally desirable. It is therefore important to characterise the behaviour and stability of these composites under the influence of freeze-thaw cycles. Water-in-oil Pickering emulsions stabilised by poly(methyl methacrylate) particles were subjected to freeze-thaw cycles of the continuous phase under two different conditions - uniform and non-uniform freezing. Confocal microscopy was used to study the emulsion behaviour and structure during these processes. The effect of droplet size and cooling rate on uniformly frozen emulsions was also considered. The final structure of the emulsion after a single freeze-thaw cycle is strongly dependent on the freezing method. Uniformly frozen emulsions show crumpled droplet structures, while non-uniformly frozen emulsions have a non-uniform structure containing foam-like regions not observed in uniform freezing. Droplet size has little effect on the final structure of uniformly frozen emulsions, which we attribute to the Laplace pressure in the droplets being orders of magnitude smaller than the pressure exerted on the droplets by the growing oil crystals. Cooling rate also has little effect as droplets become surrounded and trapped by oil crystals rapidly after samples reach the oil freezing temperature, irrespective of the speed at which they reached that temperature. When compared to surfactant-stabilised emulsions undergoing the same process, we find emulsion structure is recoverable in the surfactant case, whereas particle-stabilised emulsions are irreversibly altered.

\end{abstract}


\section{Introduction}\label{sec:intro}

Emulsions are thermodynamically metastable systems containing droplets of one immiscible liquid dispersed in another \cite{binksbookch6}. They tend to demix, a process driven by the interfacial tension $\gamma$, \textit{i.e.}~the free energy per unit area of liquid-liquid interface. Thus, they often require stabilisation by surfactants or solid particles which adsorb onto the liquid-liquid interface and limit phase separation \cite{binksbookch6, bibette,Leunissen2007PNAS}. In particle-stabilised emulsions, solid particles adsorb to the interface and stabilisation occurs due to their partial wetting by both phases, producing a trapped layer coating the interface. Instead of lowering the bare surface tension, as is the case when using surfactants, the particles provide a mechanical barrier to coalescence \cite{binksbookch6, Clegg, Binksparticles}.

The free energy of detachment, $E$, of a particle from a liquid interface is described by
\begin{equation}\label{eqn:energy}
E = \pi r^2 \gamma_{\text{ow}}(1 \pm \cos(\theta))^2 \,
\end{equation}
where $\gamma_{\text{ow}}$ is the oil-water interfacial tension, \textit{r} the radius of the particles and $\theta$ the three-phase contact angle \cite{aveyard-2}. The sign inside the brackets is positive for removing particles into the oil phase and negative for removal into the water phase. In the case of solid microparticles at a water-alkane interface, $\gamma_{\text{ow}} = 50$ mN/m \cite{hexadec-water, goebel} and $r \approx 500$ nm, so the detachment energy for $\theta \approx 90^{\circ}$ is  $\mathcal{O}(10^7)$ $k_{\text{B}}T$ (Boltzmann constant $k_{\text{B}}$ and room temperature \textit{T}) \cite{hbkchemphys}. Hence, under quiescent consitions, the particles are irreversibly adsorbed at the liquid interface, preventing coalescence \cite{melle-lask} and Ostwald ripening \cite{Tcholokova2008PCCP}. Even for particles with a much larger contact angle, \textit{e.g.}~for $\theta$ between 130 and 160$^{\circ}$ as in our system \cite{thijssenhow2011,Isa2011NatComm,Wang2016SoftMatter}, the attachment energy is still of order $10^4$ $k_{\text{B}}T$.

Particle-stabilised emulsions, commonly known as Pickering or Pickering-Ramsden emulsions, have been studied for over a century \cite{ramsden, pickering}. Over the last few decades, interest in this field has grown, mainly because they i) are model arrested systems \cite{binksbookch6} and ii) have great potential for applications in \textit{e.g.}~the agrochemical, personal-care and food industries \cite{HUNTER2008,LEALCALDERON2008}. Although there are industries where destabilising Pickering emulsions is important \cite{crude-oil-lee, binksbookch6}, retaining or enhancing the stability of Pickering emulsions is more desirable in many cases, particularly in the food industry where destabilisation of food emulsions results in damaged products or reduced shelf life \cite{degner, binksbookch8}. The characteristics of both the particles and the particle layer can affect emulsion stability; particle wettability, concentration and particle-particle interactions have all been shown to affect emulsion stability \cite{destabilise, destabilise4, aveyard-2}.

However, food products do not just sit on shelves under ambient conditions. In industry and everyday life, they are often exposed to various processes including flow, heating and freezing \cite{Tassou2010ATE}. In particular, freezing of products that employ Pickering stabilisation is important in the food industry for two reasons. Firstly, a number of products, such as ice cream \cite{Pawlik2016}, are frozen by definition and as such the effect of freezing must be understood in relation to longevity and taste. The second reason is food storage \textit{i.e.}~freezing is a useful method for delaying spoilage by microbial growth and enzyme activity \cite{George1993}. Several studies have suggested that Pickering stabilisation improves emulsion freeze-thaw stability \cite{Rayner2015,Zhu2017}, though this may not be a generic feature \cite{Ghosh2011} and/or may be caused by gelation of particles at the interface \cite{Marefati2013}. Notably, to the best of our knowledge, the microscopic understanding of how freeze-thaw processes affect Pickering emulsions is currently incomplete.

In this paper, we present a confocal microscopy study of model Pickering emulsions subjected to a freeze-thaw cycle. The emulsions consist of water-in-oil droplets, stabilised by spherical poly(methyl methacrylate) (PMMA) particles, which have been extensively used as model colloids \cite{Pusey1986}. We study the resultant structural differences between emulsions frozen under uniform and non-uniform conditions, and consider the effects of droplet size and cooling rate on the final structures. Notably, we demonstrate that freeze-thaw cycles can substantially affect the microstructure of particle-stabilised emulsions, whereas similar emulsions stabilised by surfactants rather than particles can survive freeze-thaw cycles relatively unscathed. We attribute this difference to irreversible particle adsorption, unjamming/re-jamming of interfacial particles and partial coalescence of Pickering droplets.

\section{Materials and methods}\label{sec:methods}

\subsection{Materials}\label{subsec:materials}

Poly(methyl methacrylate) particles were made according to the procedure described in Reference~\citenum{campbell}. Particles were labelled with the fluorescent dye DiIC-18 and stabilised by poly(12-hydroxystearic acid) (PHSA). The mode radius was measured by transmission electron microscopy (TEM) to be 688 nm (see Supporting Information).

Particles initially stored in decalin were transferred into \emph{n}-hexane (97\%, Fisher Scientific) through a process of centrifugation, supernatant removal, replacement with fresh \emph{n}-hexane and particle redispersal by sonication (VWR ultrasonic bath, 80 W, 45 kHz) and vortex mixing (Fisons WhirlMixer). This process was repeated five times, after which the particles were transferred into undyed \emph{n}-hexadecane by the same process.

\emph{n}-Hexadecane (ReagentPlus 99\%, Sigma-Aldrich) was filtered twice through aluminium oxide (activated, basic, Fluka) and then Nile red (technical grade, Sigma Aldrich) was dissolved into the oil at a maximum concentration of 0.7 mM; as this is diluted during sample preparation (see below), we do not expect the fluorescent dye to affect the freezing process \cite{Devodets2017arxiv07707}. The oil was then filtered once through filter paper to remove any undissolved Nile red. Distilled water was deionised using a Milli-Q (MilliPore) filtration system (to a resisitivity of at least 18 M$\Omega \cdot$cm).

\subsection{Sample preparation}\label{subsec:sampleprep}

\begin{figure}
  \includegraphics[width=0.99\textwidth]{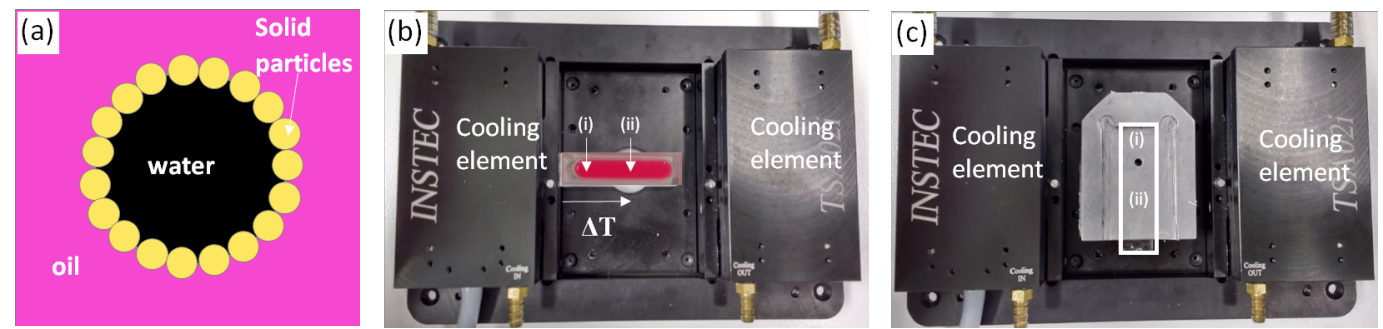}
  \caption{(a) Schematic of water droplet(black) stabilised by poly(methyl methacrylate) (PMMA) particles(yellow) in \emph{n}-hexadecane oil(magenta). (b) Temperature stage configuration for non-uniform freezing. One end of the sample cell (i) is in contact with a cooling element, while the sample cell is situated about 0.5 mm above the viewing window at (ii). (c) Temperature stage configuration for uniform freezing. The sample cell, marked with a white outline, is placed inside a metal casing and rotated 90$^{\circ}$ in relation to the non-uniform configuration to limit the temperature gradient along the sample length. In both (b) and (c), points (i) and (ii) mark the positions of the thermocouples when taking temperature measurements.}
  \label{fgr:figure1}
\end{figure}

Water-in-oil emulsions were made as follows. The particle dispersion was placed in the sonic bath for 30 minutes followed by 10 s vortex mixing. This was followed by at least a further 15 minutes in the sonic bath and 10 s vortex mixing, until the particles were dispersed and no aggregates were visible by eye. Fluorescently labelled hexadecane was added to the particle dispersion in varying volumes such that for each droplet size, the total mass of oil + particles remained constant at 2.449 g. The mixture was placed in the sonic bath for 15 minutes followed by 15 s of vortex mixing. Deionised water (2.0 g) was added and the sample was emulsified through 60 s of vortex mixing and 5 minutes standing time, repeated twice to ensure the sample was fully emulsified and the oil supernatant was clear; no excess water was observed. Figure \ref{fgr:figure1}(a) shows a schematic of the emulsion droplets using the same colours as those used in the confocal micrographs presented in Section \ref{sec:resdisc}.

The emulsion droplet diameter was controlled by the mass of particles added to the oil, and the Sauter mean droplet diameter of emulsion batches ranged between 40.4~$\upmu$m (polydispersity 40\%)and 92~$\upmu$m (polydispersity 33\%). Size was determined by confocal microscopy using a $20\times$/0.4 NA objective. Corresponding particle volume fractions in oil ranged between 6.7 and 3.0\%. The droplet diameter ($D$) is inversely proportional to the mass of particles added ($m_{\text{P}}$) and, assuming that all particles attach to the liquid-liquid interface, the two are related as follows:
\begin{equation}
\frac{1}{\left< D \right>} = \frac{s_{\text{f}}m_{\text{P}}}{6V_\text{d}} \ ,
\end{equation}
where $s_{\text{f}}$ is the specific surface area and $V_{\text{d}}$ is the volume of the dispersed phase - here water \cite{maurice, limit-co}.

Surfactant-stabilised water-in-oil emulsions were made as follows. Span 80 (S6760, Sigma-Aldrich) was used as received, added to 2.3 g of fluorescently labelled hexadecane and then subjected to 60 s vortex mixing. Deionised water (2.5 g) was added and the sample was emulsified through 60 s vortex mixing. The emulsion contained 0.9 wt-\% of Span 80 and this yielded emulsion droplets with diameters between 1 to 35 $\upmu$m, as determined by confocal microscopy.

\subsection{Freeze-thaw cycling}\label{subsec:freezethaw}

Rectangular, glass sample cells (5 mm internal path length, 1.56 ml volume, Starna Scientific Ltd) were coated with PHSA stabilising hairs to prevent the droplets from sticking to the glass. This was achieved by soaking the sample cells overnight in a solution of PHSA-Si in tetrahydrofuran (THF) and then rinsing with hexane and toluene. PHSA-Si is a comb stabiliser of PHSA chains attached to a PMMA backbone containg 5\% trimethoxysilyl propyl methacrylate and the PHSA-Si in THF solution was made in-house.

Cells were filled with 300 {$\upmu$}l of emulsion so that the emulsion did not touch the top glass lid, preventing structural changes due to capillary forces. Samples were placed inside a hotstage (Instec TSA02i with mk1000 temperature controller) and the stage temperature was recorded using the temperature controller software (WinTemp for mk1000). Iced water was pumped through the stage, while the software controlled the temperature of the sample chamber via peltier elements on either side. Two sample freezing conditions were studied - non-uniform and uniform. The experimental setup for these two cases is shown in Fig. \ref{fgr:figure1}(b) and (c) respectively.

The melting temperature of hexadecane lies in the region of 18.1 to 18.2~$^{\circ}$C \cite{hbkchemphys, hawley-condensed,hexadecaneSA2018}, therefore the stage temperature was cycled between 20~$^{\circ}$C and 10~$^{\circ}$C at selected rates between 0.5~$^{\circ}$C/min and 8~$^{\circ}$C/min. Typically, samples were held at 10~$^{\circ}$C for between 20 and 25 minutes to ensure the oil in the samples was frozen throughout before thawing. Samples were thawed by raising the temperature of the stage from 10~$^{\circ}$C back to 20~$^{\circ}$C at the same rate at which the sample was cooled. Note that at all times the temperature was held above the melting temperature of water. Each sample was subjected to a single freeze-thaw cycle.

In the case of uniform freezing, the sample cell was placed into a metal casing sealed with silicone thermal grease (RS heat sink compound plus) such that only a small part of the sample cell was not enclosed (see Fig.~\ref{fgr:figure1}(c)). This reduced the effect on the sample of the temperature gradient in the stage arising from the positioning of the cooling elements and the presence of the lower and upper glass viewing windows.

In the case of non-uniform freezing, the metal casing was not used and the sample was positioned such that one end of the sample cell was in contact with one of the cooling elements, while further along the sample cell was situated about 0.5 mm above the viewing window which acted as a heat source, the ambient temperature being higher than the stage temperature. In this way, the temperature at a given time varied with position along the sample length.

\subsection{Temperature measurements} \label{subsec:temp-measure}

\begin{figure}
  \includegraphics[width=0.99\textwidth]{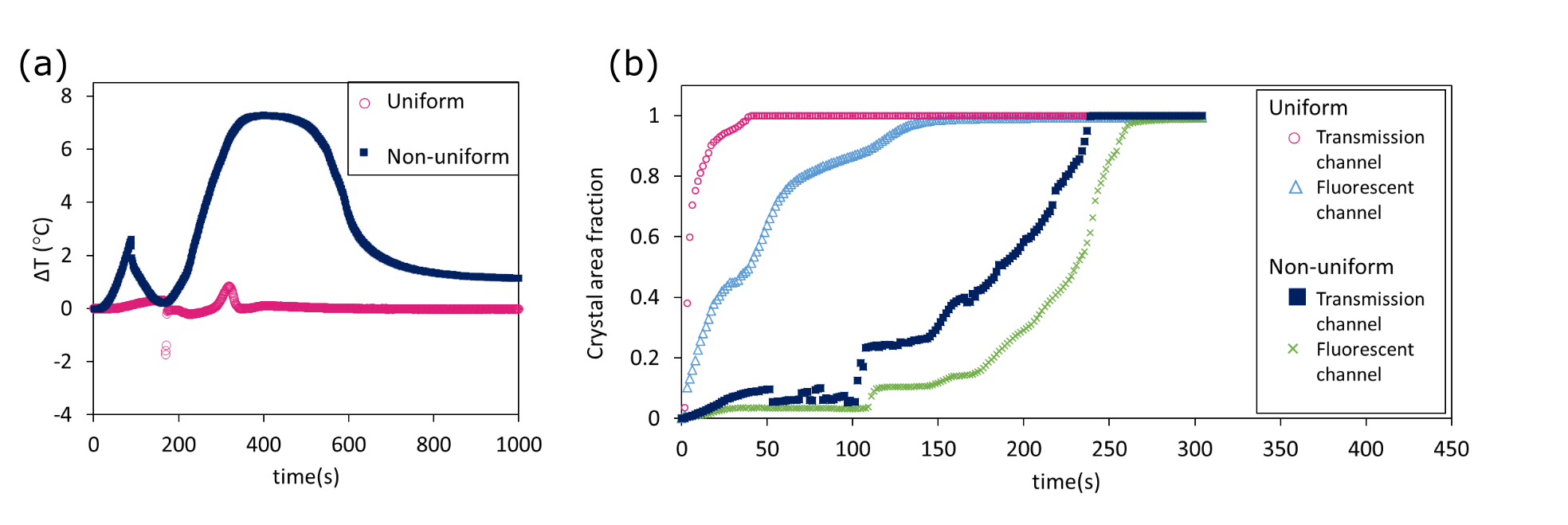}
  \caption{(a) Temperature difference between points (i) and (ii) marked in Fig.~\ref{fgr:figure1} in oil-filled sample cells during uniform(pink) and non-uniform(blue) freezing at 8~$^{\circ}$C/min. (b) Crystal fraction extracted from micrographs using image analysis for freezing of oil-filled sample cells at 2~$^{\circ}$C/min.  The crystal area fraction is measured by hand for the transmission channel and by determining  $1 - \frac{<I>}{I_0}$ for the fluorescent channel (see text for details). Some variation exists between detection channels but qualitative differences between uniform and non-uniform freezing are clear. $t=0$ is one frame (1.6 s) before crystals appear in the transmission channel.}
  \label{fgr:figure2}
\end{figure}

For each configuration, temperature versus time measurements were taken during freeze-thaw cycles at varying rates. Thermocouples (RS Pro K-type) were placed as marked in Fig.~\ref{fgr:figure1}(b-c) for each configuration and data points were recorded using a thermocouple data logger (Pico TC-08). Figure~\ref{fgr:figure2}(a) shows the temperature curves for both uniform and non-uniform cooling taken in an oil-filled cuvette. For the uniform case, there is a maximum temperature difference between the two thermocouples of less than 0.4~$^{\circ}$C for an 8~$^{\circ}$C/min cooling rate (except when the oil freezes at point (i) just before it freezes at point (ii) \textit{i.e.}~around $t = 170$~s in Fig.~\ref{fgr:figure2}(a) and a small peak around $t = 325$~s which we attribute to a residual effect of the small viewing hole in the metal casing). In the non-uniform case, there is a more pronounced temperature gradient along the sample, with a maximum temperature difference of around 7~$^{\circ}$C as shown in Fig.~\ref{fgr:figure2}(a), resulting in delayed sample freezing at point (ii) relative to point (i). These results demonstrate that the set-up in Fig.~\ref{fgr:figure1}(c) yields relatively uniform cooling, whereas the set-up in Fig.~\ref{fgr:figure1}(b) yields non-uniform cooling. This results in a difference in oil crystallization kinetics, as shown in Fig.~\ref{fgr:figure2}(b), even at cooling rates as low as 2~$^{\circ}$C/min.

Using the temperature measurements from oil-filled cells, a conversion curve can be made between the temperature recorded by the stage software and the measured temperature in the sample cell for each of the two locations. In this way, the temperature in emulsion samples during freeze-thaw cycles can be determined by noting the stage software temperature and using the conversion curve to obtain the sample temperature.

\subsection{Sample characterisation}

Samples were observed using a confocal laser scanning microscope (Zeiss LSM 700) coupled to a Zeiss Observer Z1 with an LD Plan Neofluar $20\times$/0.4 NA (Korr) objective. The 488~nm laser line was used to excite the Nile red in hexadecane and the 555~nm line to excite DiIC-18 in PMMA particles. Emission filters were used as appropriate and samples were observed in fluorescence and transmission. \par
Comparing the transmission and Nile red fluorescence channels, we note that we can also observe the freezing transition in the Nile red channel: the fluorescent signal is significantly reduced as the oil freezes. This aligns with observations of fluorescence loss upon freezing of labelled water \cite{Devodets2017arxiv07707}. This fluorescence loss could be due to rejection of dye from the oil crystals \cite{Devodets2017arxiv08510}, but we (partly) attribute it to a colour (\textit{i.e.}~spectral) shift upon freezing, based on visual observations of a 4ml Nile red-labelled hexadecane sample in a fridge.

\subsection{Image analysis}

\subsubsection{Fluorescent channel analysis}
Image stacks including both the transmission and the fluorescence channels were loaded into ImageJ \cite{ImageJ} and the oil fluorescence channel was duplicated into a separate stack. The `mean gray value' measurement was taken for each slice and recorded in a Microsoft Excel spreadsheet. As the fluorescent signal of the oil is lost upon oil crystallisation, the mean gray value decreases with increasing crystalline area in the image. The values were normalised with the first image in which only liquid oil was present and converted into a crystalline fraction by subtracting the normalised mean gray values from 1. This way, a value of 1 for $\left( 1 - \left( \left< I \right> / I_0 \right) \right)$ indicates that the sample shown in the image is fully crystalline and a value of 0 indicates that the sample is fully liquid.\\

\subsubsection{Transmission channel analysis}
Image stacks were loaded into ImageJ and the transmission channel duplicated to extract it from the multi-channel stack as above. As the background illumination is uneven, the `subtract background' process is used with an appropriate rolling ball radius to reduce the effect. The image is then thresholded to leave only the crystal showing in black. The `despeckle', `close' and `fill holes' tools are used and the canvas size is then cropped by two pixels on each side to remove the white border introduced during the process. Each image is then compared to the original and any remaining holes are filled in using the pencil and bucket fill tools. Finally, the `area fraction' measurement was taken for each slice and recorded in a Microsoft Excel spreadsheet. A value of 1 for area fraction, \textit{i.e.} a completely black image,  indicates a fully crystalline sample in the field of view, whereas a value of 0 indicates a fully liquid sample.

\subsubsection{Droplet sizing}

Particle-stabilised droplets were sized as follows: image stacks were loaded into ImageJ and the particle fluorescence channel duplicated. The clearest image of the emulsion prior to the sample freezing was extracted and then the image thresholded appropriately. As the fluorescent signal from the particles has higher signal-to-noise than the one from the oil, a `Hough transform' was applied to the image from the particle channel and the parameters varied until all the droplets were accounted for except those on the edges. This measured droplet sizes in increments of 1.25~$\mu$m. In some cases, a small number of droplets were missed and a small number of interstices included,  but not enough to significantly affect the average droplet size measured. \par

Surfactant-stabilised droplets were sized similarly, though a `smooth' function was applied twice before thresholding. `Analyze particles' was used to extract the droplet sizes, because it is a faster method than the Hough transform.

\subsection{Multiphoton characterisation}

Samples were imaged using multiphoton microscopy after undergoing a single freeze-thaw cycle. Sample cells were made by slicing off the bottom of an 8 ml glass vial and then replacing it with a microscope coverslip (Menzel-Gl{\"a}ser 22~mm $\times$ 22~mm {\#}1) . Samples were frozen using the setup described in Sec.~\ref{subsec:freezethaw}, but due to the difference in sample cells, samples were frozen in a slightly different configuration to regular samples. For non-uniform samples, the sample cells were placed such that half the sample was over the glass viewing window and half on the stage base. For uniform samples, the sample cell was too large for the metal casing, so samples were placed far from the glass viewing window so that the whole sample was close to the cooling element. This means that these samples were not imaged during the freeze-thaw process, only prior to freezing and then after thawing.\par
3D image acquisition was carried out using a custom-built LaVision TriM scope II platform with a large Luigs und Neumann inverted stage. The glass coverslip of the sample cuvette was optically coupled to a $25\times$ Nikon NA1.1 W lens (MRD 77220) with distilled water.\par
DiIC18 was excited by a pulsed Titanium:Sapphire laser (Chameleon Ultra II, Coherent) tuned to 900~nm, producing 4-8 mW in the focal plane. Nile red was excited by a Coherent/APE OPO laser tuned to 1100~nm and a laser power comparable to the Ti:Sa laser. A 740~nm LP dichroic filter was used to separate excitation from the emitted light in the direct detection port of the microscope. The non-descanned, emitted light was separated by a 560~nm long-pass beam splitter and then filtered by an ET525/70 bandpass filter for DiIC18 and an ET610 long-pass filter for Nile red. The filtered light was collected by two Hamamatsu GaAsP detectors (H7422-40-LV 5M).

\section{Results and discussion}\label{sec:resdisc}

In this Section, we consider the behaviour of particle-stabilised emulsions undergoing non-uniform or uniform cooling, mainly based on confocal microscopy observations. We explain and compare the resultant emulsion structures from the two processes. We also consider the effect of droplet size and cooling rate on the final emulsion structure for samples undergoing uniform freezing.

\subsection{Non-uniform freezing}\label{subsec:resdisc_nonuniform}

\begin{figure}
  \includegraphics[width=0.99\textwidth]{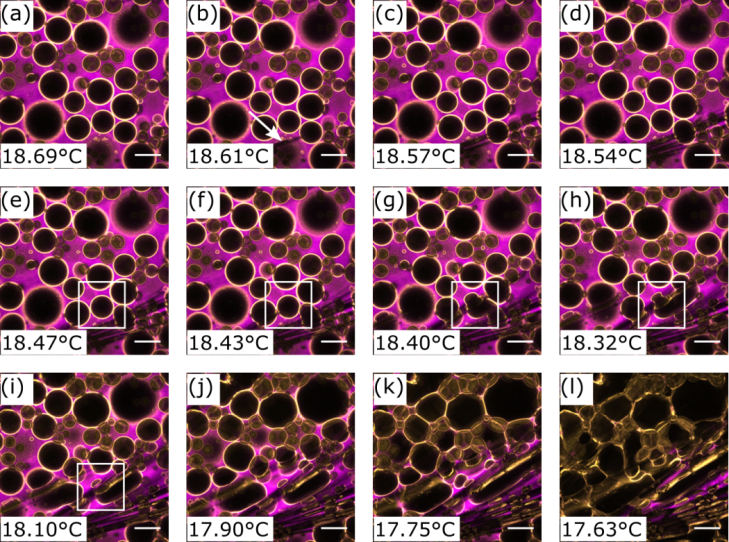}
  \caption{Confocal micrographs of a particle(yellow)-stabilised water-in-oil(magenta) emulsion taken during an 8~$^{\circ}$C/min non-uniform freeze. Temperatures shown are those determined close to the field of view (see Sec.~\ref{subsec:temp-measure} for method). Images are taken (a-h) 3.1~s and (h-l) 15.7~s apart. Note that the fluorescent signal from the oil is lost as it crystallises (see text for details). Oil crystals begin to form in (b) and droplets are squeezed together into foam-like structures as shown in (j-l). The white rectangle highlights droplet splitting. Scale bars: 100 $\upmu$m.}
  \label{fgr:figure3}
\end{figure}

We first consider non-uniform freezing of emulsions using the setup shown in Fig.~\ref{fgr:figure1}(b). The sample starts freezing from the end nearest to point (i) and is imaged at point (ii). Figure~\ref{fgr:figure3} shows a time series of images of an emulsion undergoing non-uniform freezing. Particle~(yellow)-coated water droplets are initially spherical and are dispersed throughout the oil~(magenta). As the temperature in the cooler region of the sample drops below the oil freezing temperature, dark streaks appear in the oil near the arrow in Fig.~\ref{fgr:figure3}(b) and grow over time. Water droplets move away from the newly formed dark regions and move closer to each other and in Fig.~\ref{fgr:figure3}(j-l) we see the formation of a region with foam-like structure \cite{maurice, cellular, zou-foam, liquid-foam}.

The appearance of dark regions in the oil suggests that the oil is crystallising, which is in line with the temperature being below the melting point of the oil. The crystals can also be observed in the transmission images (not included here). Crystals form first in the cooler part of the sample and grow into the warmer, still liquid, regions of the sample. Some oil crystals appear to skewer water droplets, seen as droplet shapes being distorted in the crystal-growing region (as in Fig.~\ref{fgr:figure3}(d) lower right corner) and in Fig.~\ref{fgr:figure3}(e)-(i) in the white box we see a single droplet splitting and becoming two particle-coated droplets.\par
Oil crystals also appear to push droplets out of the way. As the crystals form, the volume of liquid in which the droplets can move decreases, causing droplets to move closer together and exclude liquid oil from between them as they move. The liquid region can be further reduced when crystals grow at an acute angle to one another, cutting off droplets from accessing other liquid regions of the sample. Eventually, the droplets are packed together in a foam-like structure as this is the most compact structure for the droplets to form without breaking their particle coating; this structure remains after the oil thaws. This behaviour is reminiscent of particle-stabilised emulsions retaining a foam-like structure after having been compressed by centrifugation \cite{maurice}.\par
The foam-like regions in the sample generally form in the region of the sample situated above the temperature stage viewing window. This is the warmest part of the sample, as the glass window is in direct contact with the (ambient) air outside the stage. In this region, the oil will be liquid for the longest amount of time, thus it is the easiest place for droplets to collect as the oil freezes elsewhere. The size of the foam-like region varies with sample because it is strongly dependent on the oil crystal growth pattern, a parameter over which we currently have limited control.\par
\begin{figure}
  \includegraphics[width=0.49\textwidth]{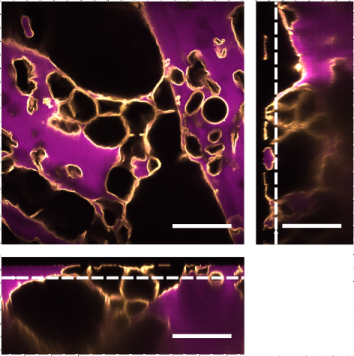}
  \caption{3D structure of water-in-oil~(magenta) emulsion stabilised by PMMA particles~(yellow) after undergoing a single non-uniform freeze-thaw cycle: $xy$ (square), $xz$ and $yz$-slices are shown with the $xy$ slice taken at the height shown by the white dashed lines; cooling rate during freezing stage: 4~$^{\circ}$C/min. Scale bars: 100 $\upmu$m.}
  \label{fgr:figure4a}
\end{figure}

So far, we have considered only the 2-dimensional structure of these samples. Using multiphoton imaging we can image further into the samples, allowing us to observe whether or not the structure persists in the $z$-direction. Figure~\ref{fgr:figure4a} shows $xz$ and $yz$ micrographs of a non-uniform sample. We see that that the foam-like structure persists for at least two layers in the $z$-direction, showing that this is not purely a wall effect. Although we cannot image at single-particle resolution due to the constraints of the two-photon setup and the sample cell requirements, comparing the thickness of the particle assembly between droplets that have been squeezed together with the thickness of the particle coating of a single droplet in Fig. \ref{fgr:figure4a}, it appears that 'zipping' i.e. sharing of particles between droplets does typically not occur.

\begin{figure}
  \includegraphics[width=0.99\textwidth]{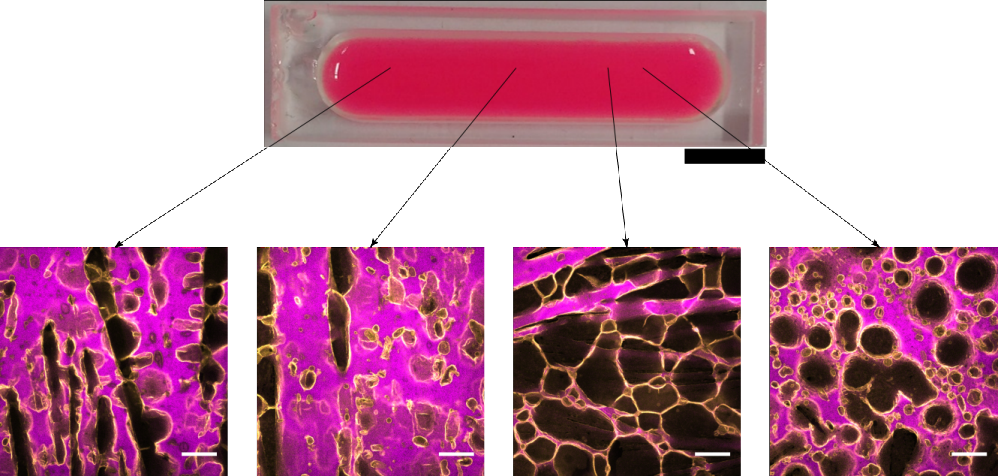}
  \caption{Series of confocal micrographs taken along the length of a water-in-oil~(magenta) emulsion stabilised by PMMA particles~(yellow) that was first cooled under non-uniform conditions and then thawed; initial droplet diameter was 92~$\upmu$m and cooling rate was 8~$^{\circ}$C/min. Different regions of the sample display different structures due to the non-uniform cooling. Arrows show the approximate position at which the images were taken. Scale bars: 2~mm on sample cell image and 100~$\upmu$m on emulsion images.}
  \label{fgr:figure4}
\end{figure}

As one might expect, non-uniform cooling of the sample leads to a non-uniform structure of the final thawed sample. This can be seen in Fig.~\ref{fgr:figure4}, which shows images of an emulsion sample after a single freeze-thaw cycle. We see a foam-like structure, coalesced droplets and individual droplets all in the same emulsion. As the crystal growth is not uniform throughout the sample, droplets are affected differently in different regions of the sample, leading to the non-uniform structure observed.

Even though non-uniform cooling may well be more relevant for certain applications, \textit{e.g.}~putting a food product in a freezer, it presents challenges when attempting to extract trends due to the non-uniformity of the resulting samples. Hence, in the remainder of this section, we focus on results for uniform freezing.

\subsection{Uniform freezing}\label{subsec:resdisc_uniform}

\begin{figure}
  \includegraphics[width=0.99\textwidth]{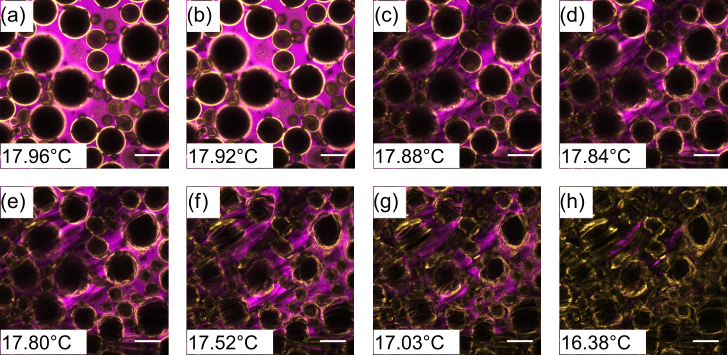}
  \caption{Confocal images of a water-in-oil~(magenta) emulsion stabilised by PMMA particles~(yellow) taken during uniform freezing at 8~$^{\circ}$C/min. Temperatures shown are those determined close to the field of view (see Sec.~\ref{subsec:temp-measure} for method). Images are taken (a-e) 3.1 s and (e-h) 15.7 s apart. As the oil crystallises, it fills the field of view within a short space of time in comparison to non-uniform freezing (Figure \ref{fgr:figure3}). Scale bars: 100 $\upmu$m.}
  \label{fgr:figure5}
\end{figure}

Having considered non-uniform freezing, we now consider samples undergoing uniform freezing in the setup shown in Fig.~\ref{fgr:figure1}(c). Minimizing the temperature gradient means that all the oil in the sample freezes at the same time. Figure~\ref{fgr:figure5} shows a time series of images of an emulsion undergoing uniform freezing. As for the non-uniform case, the water droplets begin spherical and dispersed throughout the oil. Once the temperature in the sample drops below the oil freezing temperature, dark streaks appear across the whole image (see Fig.~\ref{fgr:figure5}(c)) and then over time the fluorescent signal from the oil is completely lost. In contrast to the non-uniform case, water droplets do not move closer together but become crumpled, shown by the fluorescence of the particles which are stabilising the droplets.

As the temperature gradient in the sample is minimal, the oil begins to freeze across the whole field of view (640 $\upmu$m $\times$ 640 $\upmu$m) at effectively the same time. Droplets are surrounded by oil crystals within the time between two subsequent frames (taken 3.1 s apart), effectively pinning them. As the crystals grow, the volume available to the droplet decreases and the shape of that cavity is defined by the surrounding crystals. As the water is still liquid, the droplet crumples to take on the shape of the cavity. Intriguingly, droplets retain their crumpled shape upon thawing, \textit{i.e.}~they do not fully relax to a spherical shape. We have tested the stability of this structure for a period of three days and no changes in structure were observed.\par
We can explain these observations in terms of unjamming  and rejamming. As the sphere has the smallest area for a given volume, the crumpling leads to an increase in droplet area. As the particles are irreversibly attached to the liquid interface and there are no free particles in the continuous phase to fill up any extra interface, they are no longer closely packed on the droplet surface. This unjamming allows the particles to move as they are released from trapping by the oil crystals upon thawing. The droplet can therefore undergo minor changes in shape until the interfacial particles are again jammed. However, the particles may rejam at a packing fraction below close packing \cite{onoda}, impeding the droplet from returning to a spherical shape.\par
\begin{figure}
  \includegraphics[width=0.49\textwidth]{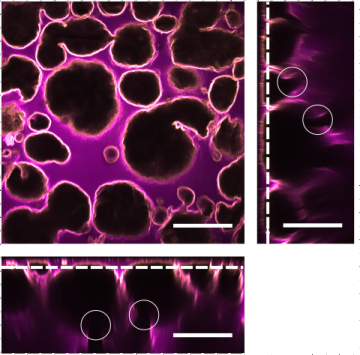}
  \caption{3D structure of water-in-oil~(magenta) emulsion stabilised by PMMA particles~(yellow) after undergoing a single uniform freeze-thaw cycle: $xy$ (square), $xz$ and $yz$-slices are shown with the $xy$ slice taken at the height shown by the white dashed lines; cooling rate during freezing stage: 4~$^{\circ}$C/min. White circles denote locations where droplets have partially coalesced. Scale bars: 100 $\upmu$m.}
  \label{fgr:figure7}
\end{figure}

We can again image further into the samples using multiphoton imaging, allowing us to observe structure in the $z$-direction. Figure~\ref{fgr:figure7} shows $xz$ and $yz$ micrographs of a uniform sample. We see that the droplets have a non-spherical profile and some appear dumbbell-shaped. This is an indication of partial coalescence between droplets. Relaxation to a spherical shape after coalescence is prevented by the fact that the total surface area of two single droplets is larger than the surface area of a single droplet at fixed volume ($A / V \propto 1/D$). Hence, the particles jam the interface once they completely cover the surface of the partially-coalesced droplet. This partial coalescence also leads to interfacial particles rejamming before a spherical shape is reached.\par
\begin{figure}
  \includegraphics[width=0.99\textwidth]{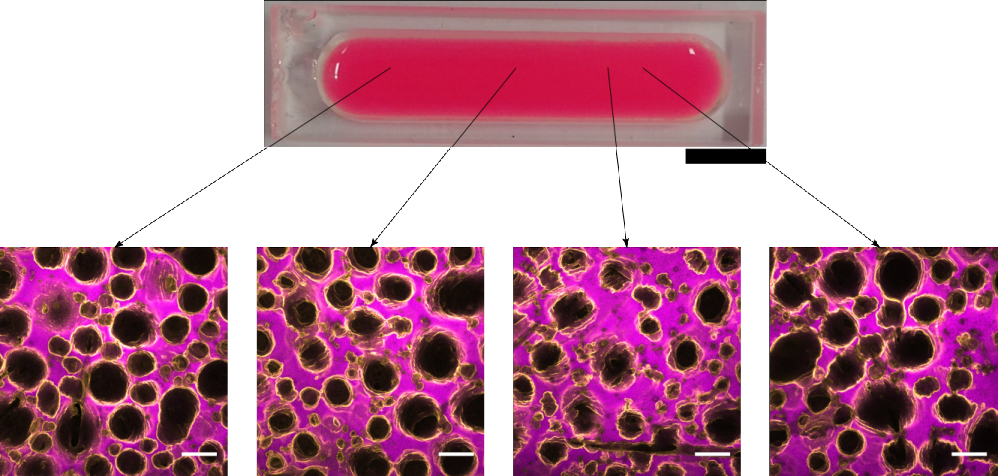}
  \caption{Series of confocal micrographs taken along the length of a water-in-oil~(magenta) emulsion stabilised by PMMA particles~(yellow) that was first cooled under uniform conditions and then thawed; initial droplet diameter was 92~$\upmu$m and cooling rate was 8~$^{\circ}$C/min. Arrows show the approximate locations at which each image was recorded. In each image, a mixture of individual and partially coalesced droplets can be seen. Scale bars: 2~mm on sample cell image and 100~$\upmu$m on emulsion images.}
  \label{fgr:figure6}
\end{figure}

Given uniform cooling, we expected a uniform structure after a freeze-thaw cycle. Indeed, Fig.~\ref{fgr:figure6} shows images of an emulsion at different points along the sample length after undergoing a single uniform freeze-thaw cycle. In each image, we see crumpled droplets along with some droplets which are connected to their neighbours via narrow, particle-stabilised `necks'. These micrographs look qualitatively similar regardless of where along the sample they have been recorded, in complete contrast to the behaviour observed in the non-uniform case (see Fig.~\ref{fgr:figure4}).\par

\subsection{Uniform freezing - droplet size and cooling rate}

\begin{figure}
  \includegraphics[width=0.99\textwidth]{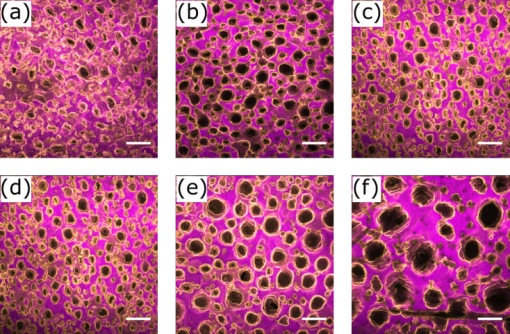}
  \caption{(a-c) Final thawed emulsion structures for $47$~$\upmu$m water droplets in oil~(magenta), stabilised by PMMA particles~(yellow), after first having been frozen at rates of (a) 0.5~$^{\circ}$C/min, (b) 4~$^{\circ}$C/min and (c) 8~$^{\circ}$C/min; the qualitative behaviour of the samples is the same irrespective of the cooling rate. (d-f) Final thawed emulsion structure for similar emulsions after first having been frozen uniformly at 8~$^{\circ}$C/min; initial droplet diameters were (d) 47 $\upmu$m, (e) 69 $\upmu$m and (f) 92 $\upmu$m. In each case, crumpled droplets are observed and the samples qualitatively have the same structure aside from the differing droplet size. Scale bars 100 $\upmu$m.}
  \label{fgr:figure8}
\end{figure}

Having established that samples submitted to a uniform freeze-thaw cycle have a qualitatively uniform structure, we can explore the effects of varying droplet size and cooling rate on the morphology. Figures~\ref{fgr:figure8}(a-c) show emulsions with 47~$\upmu$m diameter droplets after undergoing a single freeze-thaw cycle at increasing cooling rates. In each case we again see qualitatively the same behaviour and there is no clear trend in structure as cooling rate is increased. As the final structure of the sample is determined only from the point of crystallisation onwards, the cooling rate, which primarily determines how quickly the sample reaches the oil freezing temperature, has little effect on the final sample morphology in the range of rates considered here.\par
Having considered cooling rate, Figs.~\ref{fgr:figure8}(d-f) show emulsions of increasing droplet size after a single freeze-thaw cycle at a fixed cooling rate of 8~$^{\circ}$C/min. In all cases, crumpled droplets are present, with some connected to neighbours as seen earlier (see Fig.~\ref{fgr:figure6}). Bar the droplet size, the morphology is qualitatively the same in each emulsion and there is no clear structural trend. To explain this, we consider the Laplace pressure, $P_{\text{L}}$, which, for a spherical droplet of diameter $D$, is given by:
\begin{equation}
P_{\text{L}} = \frac{4\gamma}{D} \ ,
\end{equation}
where $\gamma$ is the droplet surface tension. Taking a maximum droplet diameter of 92~$\upmu$m and a minimum of 40~$\upmu$m, with surface tension of order 50 mN m$^{-1}$ \cite{hexadec-water, goebel}, gives a value for $P_{\text{L}}$ between 2.2 and 5.0 kPa. The Laplace pressure opposes the pressure applied to the droplets from the growing oil crystals. As we observe that the effect of changing the droplet size is small, we surmise that $P_{\text{L}}$ is small in comparison to the oil crystallisation pressure. This aligns well with measurements of the pressure exerted on an object by growing ice crystals (of order 20 kPa) \cite{Connell1971Glaciology}. Moreover, note that the confocal micrographs indicate that the oil-crystal size is smaller than the smallest droplet.

\subsection{Comparison with surfactant-stabilised emulsions}\label{subsec:resdisc_surfactant}

To understand the role of the interfacial particles in emulsion behaviour and microstructure upon freeze-thaw cycling, we have also considered the freezing of a surfactant-stabilised emulsion. The Laplace pressure is similar for both surfactant and particle-stablised emulsions used in these experiments, as both the surface tension and the droplet size are smaller in the surfactant-stabilized ones. Hence, if samples are otherwise treated in the same manner, any differences in behaviour between surfactant and particle-stabilised emulsions can be attributed to the interfacial particles. It should be noted that the water droplets in these experiments are significantly smaller than those in the emulsions presented so far, being on average around 11~$\upmu$m rather than 40 to 92~$\upmu$m in diameter, but we have shown above that droplet size does not affect the structure of our particle-stabilised emulsions undergoing uniform freezing.\par
\begin{figure}
  \includegraphics[width=0.99\textwidth]{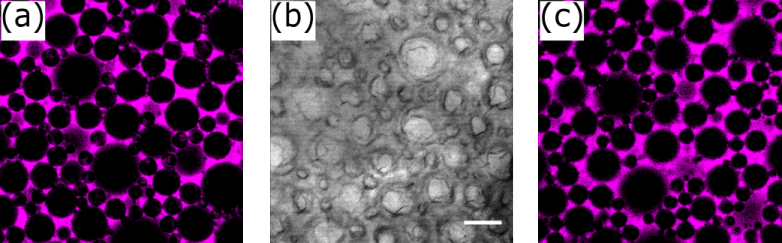}
  \caption{Confocal images of a surfactant (Span 80) stabilised water-in-oil~(magenta) emulsion undergoing uniform cooling at 4~$^{\circ}$C/min. (a) Initial emulsion in fluorescence channel and (b) frozen emulsion as imaged in transmission channel (see text for details); the droplets have crumpled in a manner qualitatively similar to particle-stabilised emulsions. (c) Thawed emulsion in fluorescence channel; the droplets have recovered their spherical shape. Scale bars: 25 $\upmu$m.}
  \label{fgr:figure9}
\end{figure}

Figure~\ref{fgr:figure9} shows a surfactant-stabilised emulsion before freezing, while frozen and after thawing. The oil is magenta as before, but now this is the only fluorescent dye in the sample.  Note that due to the weakened fluorescence signal of the oil after freezing, the image for the frozen sample is taken from the transmission channel. Initially, the water droplets are spherical and appear black. Once frozen, the droplets show similar crumpling behaviour to that of the particle-stabilised emulsions but, upon thawing, the droplets return to a spherical shape. The average droplet diameter before freezing is 11.4(8) $\upmu$m and after thawing is 10.2(6)~$\upmu$m. The small difference can be explained by small droplet movements relative to the focal plane upon freeze-thaw cycling.\par
We have shown above that the microstructure of particle-stabilised emulsions can be changed irreversibly by freeze-thaw cycling, whereas surfactant-stabilised emulsions appear similar before and after freeze-thaw cycling. We attribute this difference to the irreversible adsorption of particles at liquid interfaces (see Eq.~(\ref{eqn:energy})), whereas surfactants can adsorb and desorb at relatively fast timescales \cite{Binksparticles}. Hence, if a surfactant-stabilised droplet gets deformed (freezing), additional surfactant from nearby bulk phase that is still liquid can adsorb onto the droplet surface to stabilise the additional liquid interface. If the constraints on the droplet shape are removed (thawing), some of the surfactant molecules can detach from the droplet surface, if necessary, to allow relaxation to a spherical shape; this is because surfactants are smaller in size and correspondingly have a lower detachment energy (equation \ref{eqn:energy}).\par
Consider a droplet stabilised by particles that are closely packed at its surface ($\varphi_{\mathrm{2D}} \approx 0.907$ for monodisperse discs \cite{Meyer2010}). Upon deformation (freezing), the droplet surface area increases (the spherical shape has the lowest area per unit volume), which may lead to unjamming of interfacial particles. If the constraints on droplet shape are removed (thawing), the droplet may partially relax, but the interfacial particles may jam at a lower surface fraction, \textit{e.g.}~random close packing $\varphi \approx 0.89$ or random loose packing $\varphi \approx 0.84$ \cite{Meyer2010}, which means the droplet cannot relax to a spherical shape. Calculations of specific surface area for droplets prior to freezing show that the particles are close to hexagonal packing, thus the particles cannot jam at a higher packing fraction but can only jam at a lower one.\par
Van Hooghten \textit{et al.}~have shown that in expansion-compression cycles of a particle-coated interface in a Langmuir trough, successive cycles show a decrease in the area at which the surface pressure increases rapidly, suggesting that the particle packing fraction increases with each cycle \cite{vHooghten}. As the emulsification of a particle-stabilised emulsion via vortex mixing is an energetic process in which droplets without a full coverage of particles will coalesce with other droplets until full coverage is reached, we effectively start with a highly packed particle coating on the droplets. Upon freezing, the droplet surface expands, analogous to a Langmuir trough expansion cycle, and when it thaws, the compression is far less energetic than during emulsification, thus we compress to a lower packing fraction. The effect of rejamming at lower surface fraction is exacerbated by partial coalescence, which leads to a reduction in interfacial area $A$ per unit volume $V$. Specifically, $A/V \propto L^{-1}$ with $L$ the typical size of the (partially coalesced) droplet: partial coalescence leads to an increase in $L$ and hence a decrease in available surface area as volume is conserved.\par
It could be argued that the difference in behaviour between surfactant and particle-stabilised emulsions is due to the lower interfacial tension of surfactant-stabilised droplets. However, it has been shown that under a given applied osmotic pressure, normalised by the (reduced) interfacial tension and droplet radius, surfactant-stabilised emulsions compress to a higher packing fraction of droplets than particle-stabilised emulsions \cite{maurice}. Hence, if one considers the reduction of interfacial tension by surfactants only, it is in fact surprising that the surfactant-stabilized droplets outperform the particle-stabilised ones during freeze-thaw. This implies that the difference in behaviour that we have observed between surfactant and particle-stabilized emulsions is due to the presence of the particles.\par
Considering the relevance to food products, freezing is a widespread method to preserve food products and to facilitate their transport. Pickering stabilisation is also crucial for, or can be used to enhance, shelf life of food products. However, our results indicate that freezing particle-stabilised emulsions can lead to irreversible changes in microstructure, especially in the case of non-uniform freezing (which is more likely in production and use). Hence, even though Pickering stabilisation has been suggested for enhancing the shelf life of food products, if they are likely to be frozen in production and/or use, these food products may not necessarily benefit from Pickering stabilisation. In some cases, the irreversible damage may be limited, \textit{e.g.}~in those cases where the droplet phase freezes first, but this may not always be a viable strategy.

\section{Conclusions}\label{sec:conclusions}

We have used confocal microscopy to investigate the microstructure of water-in-oil Pickering emulsions undergoing freeze-thaw cycles. We have investigated the behaviour of samples undergoing both uniform and non-uniform freezing, as well as considering the effects of altering the droplet size and cooling rate.\par
Non-uniform freezing of emulsions produces non-uniform post-thaw samples which display several different structures, notably foam-like regions which are a result of droplets being pushed together by growing oil crystals. During uniform freezing, all areas of the sample crystallise effectively at the same time, meaning droplets are unable to move. Instead, they crumple as they are forced into non-spherical cavities within the frozen oil. Deformed droplets do not relax to spherical upon thawing, which we attribute to irreversible particle adsorption: interfacial particles unjam when the droplet is deformed (freezing) and rejam when the restrictions on droplet shape are removed (thawing). However, they may rejam at a lower surface fraction, leading to a non-spherical droplet shape after freeze-thaw cycling; this is exacerbated by coarsening via partial coalescence, which leads to a smaller area available for interfacial particles for a given volume of dispersed phase.\par
Finally, we show that both cooling rate and droplet size have no significant effect on the final emulsion microstructure. The latter also means we can compare our results to similar experiments employing surfactant-stabilised emulsions that have a relatively small droplet size. We show that these surfactant-stabilised emulsions survive freeze-thaw cycles relatively unscathed. We argue that these results are relevant in the context of food products: Pickering stabilisation can be used to enhance their shelf life, but it can also result in irreversible changes to the microstructure if those products are subjected to freeze-thaw cycling during processing or use.

\begin{acknowledgement}

We thank A.B.~Schofield for synthesising the particles. K.L.D.~acknowledges studentship funding from EPSRC with the Condensed Matter Doctoral Training Centre (CM-CDT) under grant number EP/L015110/1. J.H.J.T.~acknowledges The University of Edinburgh for a Chancellor's Fellowship. We thank the Confocal and Advanced Light Microscopy (CALM) Facility for equipment provision and technical support.

\end{acknowledgement}

\begin{suppinfo}

The following files are available free of charge.
\begin{itemize}
  \item Filename: \emph{Supporting-information.pdf}, containing additional information on particle-size distribution.
\end{itemize}

\end{suppinfo}

\bibliography{achemso-demo.bbl}

\end{document}